\documentclass[aps,pre,groupedaddress]{revtex4}
\usepackage{graphicx}
\usepackage{color}

\begin{document}

\title{Dynamics of Molecular Motors with Finite Processivity on Heterogeneous Tracks}

\author{Yariv Kafri$^*$, David K. Lubensky$^\#$ and David R. Nelson$^*$}
\affiliation{$^*$ Department of Physics, Harvard University,
Cambridge, MA 02138} \affiliation{$^\#$ BioMaPS Institute, Rutgers
University, Piscataway NJ 08854 and Bell Labs, Lucent
Technologies, Murray Hill, NJ 07974}

\date{\today}

\begin{abstract}
The dynamics of molecular motors which occasionally detach from a
heterogeneous track like DNA or RNA is considered. Motivated by
recent single molecule experiments, we study a simple model for a
motor moving along a disordered track using chemical energy while
an external force opposes its motion. The motors also have finite
processivity, i.e., they can leave the track with a position
dependent rate. We show that the response of the system to
disorder in the hopping off rate depends on the value of the
external force. For most values of the external force, strong
disorder causes the motors which survive for long times on the
track to be localized at preferred positions. However, near the
stall force, localization occurs for {\it any} amount of disorder.
Existence of localized states near the top of the band implies a
stretched exponential contribution to the decay of the survival
probability. To obtain these results, we study the complex
eigenvalue spectrum of the time evolution operator. A similar
spectral analysis also provides a very efficient method for
studying the dynamics of motors with infinite processivity.
\end{abstract}


\maketitle

\section{Introduction}

Single molecule experiments which study molecular motors provide a
powerful tool for understanding their function
\cite{Bustamante2003}. By applying a mechanical force, one can often
discern details of their reaction steps. For example, a recent paper
considered the motion of a kinesin moving along a microtubule under
the influence of a force opposing its motion \cite{Visscher99}. By
measuring the relationship between the applied force, $F$, and the
velocity of the kinesin, $v$, at different ATP concentrations it was
possible to infer that the chemical reaction cycle contains at least
one load dependent transition.

More recently, it has been possible to study, using similar
techniques, the motion of RNA polymerase (RNAp)
\cite{Davenport00,Wang98}, DNA polymerase (DNAp) \cite{Wuite00},
helicases \cite{Ha2002} and $\lambda$-exonuclease
\cite{Perkins2003}. In contrast to kinesin and myosin, which move
along homogeneous polymer filaments, these motors move along DNA,
which is inherently a heterogeneous track, with the energy landscape
determined by the nucleotide sequence \cite{RNAp}. Indeed, the
dynamics of these motors seems far richer than that of kinesin
\cite{Visscher99} or, say, myosin V \cite{Yildiz2003}, which move
along periodic filaments. Theoretically, it is expected that
heterogeneous and homogeneous (or periodic) tracks can give rise to
very different motor dynamics.  On long time and large length scales
the motion of motors moving along a homogeneous track is described
well by a random walker moving along a tilted potential. In contrast,
the motion of molecular motors which use chemical energy to move along
disordered filaments is described by a random walker moving on a {\it
random forcing} energy landscape \cite{Harms97,Kafri04}. The effective
energy difference between two points separated by $m$ nucleotides
scales as $\sqrt{m}$. The large energy barriers implied by such
landscapes lead to anomalous dynamics \cite{Bouchaud90} when the
overall tilt of the energy landscape is small; the displacement of the
motor grows as a sublinear power of time. For molecular motors this
corresponds to applying an external force strong enough to place the
system near the stall point of the motor.

The theories described above do not treat the effect on the dynamics
of a (position-dependent) detachment probability of motors from the
track. We will refer to models where no unbinding of the motor from
the track is allowed as ``infinite processivity'' models. In this
paper we study theoretically the effect of a finite processivity on
the behavior of molecular motors. Previous work has considered the
effect of detachment mainly on the dynamics of many molecular motors
or with homogeneous tracks
\cite{Leibler,Ajdari,Lipowsky01,Frey,Prost}. We consider a motor which
can leave the polymer on which it is moving but can never rebind to
it. This case is relevant to single molecule experiments which follow
a specific motor or a dilute concentration of motors on a long DNA
track. In such experiments (especially if a background hydrodynamic
flow washes away detached motors) the probability of rebinding to the
track after detachment is negligible.

\begin{figure}
\includegraphics[width=7cm]{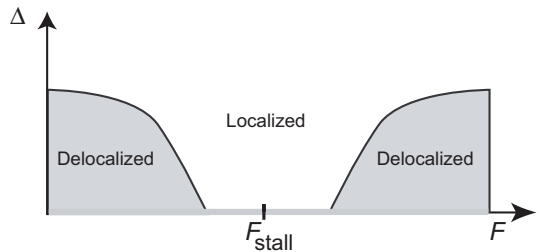} \caption{Schematic behavior of the
influence of disorder in the rates for leaving the track on the
dynamics of the motors. $\Delta$ represents the strength of the
disorder, measured through the variance of the hopping off rates
divided by the square of their mean. Note that when $\Delta=0$ the
motors are always delocalized except at $F_{\rm stall}$. This is
emphasized by the shading. It should be stressed that the notions
``localized'' and ``delocalized'' here refer to motors which
remain on the track for long times.\label{fig:dphase}}
\end{figure}

We show that when molecular motors are moving along a homogeneous
(or periodic) filament the long time dynamics of those motors that
remain attached is unaltered by detachment events. However, for
motors moving along a heterogeneous track this is not the case.
For proteins such as RNAp and DNAp that walk along DNA, the
detachment rate depends on the monomer on which the motor is
located. The disorder in the detachment rates is thus correlated
with the disorder in the hopping rates as both are determined by
the same DNA sequence. When the detachment rate varies from
monomer to monomer, the motors' dynamics can be influenced by
spatially localized eigenfunctions of the evolution operator. For
the regime where the displacement of the motor is linear as a
function of time, strong disorder causes the last motors which
remain attached to the track to stall out before falling off.
Moreover, when the displacement of the motor as a function of time
is sublinear, {\it any} amount of disorder in the hopping off
rates causes the last motors which remain attached to the track to
halt before falling off. A schematic representation of the
resulting ``phase diagram'' for the dynamics is shown in Fig.
\ref{fig:dphase}.

To show these results we use methods developed in
\cite{HatanoNelson96,HatanoNelson97}, in the context of the
physics of vortices, and numerically study the eigenvalue spectrum
of the non-hermitian evolution operator for the probability
distribution of the motor. As will be shown, the characteristics
of this operator's spectrum (which include localized and
delocalized states as well as a mobility edge) allow the long
time, large length-scale dynamics to be obtained in a
straightforward manner. Interestingly, the {\it dynamics} of
motors on heterogeneous tracks with finite processivity is similar
to the non-hermitian {\it statistical mechanics} arising in models
of vortex physics \cite{HatanoNelson96,HatanoNelson97}.

To support these results further we also study analytically a toy
model which consists of a directed walker among traps with a broad
distribution of release times. The model was studied for the case
of infinite processivity in \cite{Bouchaud90}. We show that in the
infinite processivity case the model yields the same spectrum as
observed in the numerics. Moreover, we show that for a motor with
finite processivity, when the corresponding infinite processivity
model shows sublinear drift, any amount of disorder in the hopping
off rates leads to localization of the eigenfunction, consistent
with our numerical results.

The paper is organized as follows: In Section II the model we
study is introduced and known results for the infinite
processivity limit are reviewed. In Section III the infinite
processivity model is analyzed using the eigenvalue spectrum of
the evolution operator. It is shown that the long time, large
length-scale behavior of the model can be extracted from the
properties of the spectrum. In Section IV we study the spectral
properties of an evolution operator that takes account of
detachment of motors from the track and their physical
implications. Specifically, the implications for measurements of
the probability of finding a motor on the track as a function of
time are discussed. Finally, in Section V the analysis of the toy
model is presented.

\section{The Model}
\begin{figure}
\includegraphics[width=8cm]{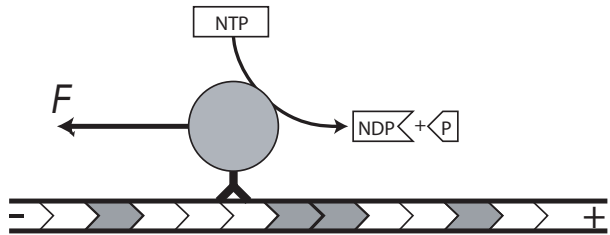} \caption{Setup modelled. The motor is
moving from the $-$ end to the $+$ end, driven by hydrolysis of
nucleotide tri-phosphate. A force is pulling on the motor in the
opposite direction. The track shown is made up of two types of
monomers (depicted as shaded and white areas). Although this
schematic suggests a microtubule with disorder in the protein
constituents, we actually have in mind motors moving on DNA or RNA
templates, with four distinct nucleotides. Elaborations which make
our model more realistic at a microscopic level should not affect
the predictions for long time, large length-scale
dynamics.\label{fig:motorsu}}
\end{figure}

\begin{figure}
\includegraphics[width=8cm]{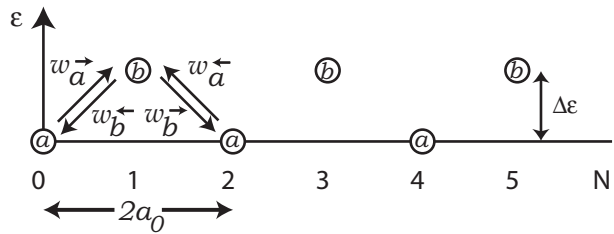} \caption{Graphical representation
of the model for molecular motors. The distinct even and odd sites
are denoted by $a$ and $b$ respectively. \label{fig:motor}}
\end{figure}

The model we consider was introduced and studied in the limit of
infinite processivity in \cite{Kafri04}. It is inspired by
previous models of molecular motors
\cite{Fisher99,Kolo00,Harms97,Prost94,Julicher97}, but is simple
enough so that exact solutions can be found, in the limit of
infinite processivity, with and without disorder. The model is
defined on a discrete lattice, $x=0,1,2 \ldots$, with distinct $a$
(even) and $b$ (odd) sites and a distance $a_0$ between lattice
points. A monomer of the track (a nucleotide, say) is taken to be
of size $2a_0$. The arrangement is shown schematically in Figs.
\ref{fig:motorsu} and \ref{fig:motor}. To model the two internal
states of the model we take even sites to have an energy
$\varepsilon=0$ while odd sites have an energy $\varepsilon=\Delta
\varepsilon$. The transition rates depicted in Fig.
\ref{fig:motor} take the form
\begin{eqnarray}
w_a^\rightarrow&=&(\alpha e^{\Delta \mu /T} + \omega)e^{-\Delta
\varepsilon/T-f/2T}
 \nonumber \\
w_b^\leftarrow&=&(\alpha  + \omega)e^{f/2T}
 \nonumber \\
w_a^\leftarrow &=&(\alpha' e^{\Delta \mu /T} + \omega')e^{-\Delta
\varepsilon/T+f/2T} \label{eq:rates}
\\
w_b^\rightarrow&=&(\alpha'  + \omega')e^{-f/2T} \;,  \nonumber
\end{eqnarray}
where we have set the Boltzmann constant to be $k_B=1$. Note that
there are two parallel channels for the transitions. The first,
represented by contributions containing $\alpha$ and $\alpha'$,
arise from utilization of the chemical energy $\Delta \mu$. The
chemical energy difference could be, for example, a result of an
excess concentration of NTP's (or just ATP) with respect to its
thermal equilibrium value. The second channel, represented by the
terms containing $\omega$ and $\omega'$, correspond to thermal
transitions unassisted by chemical energy. In addition, the
externally applied force $F$, with $f=Fa_0$, biases the motion. To
model finite processivity we add rates $w_{\rm off}^a$ and $w_{\rm
off}^b$ corresponding to the detachment from the track at even and
odd sites respectively.

\begin{figure}
\includegraphics[width=8cm]{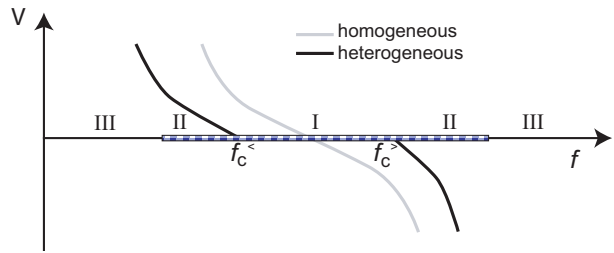} \caption{Schematic
behavior of the velocity for a heterogeneous linear motor track as
a function of the applied force. It is assumed that chemical
forces (from the NTP hydrolysis) lead to a positive velocity in
the absence of a force. The anomalous dynamics arises in the
vicinity of a stall force defined by Eq. \ref{eq:sinailoc}. The
different dynamical regimes defined in the text are denoted in the
figure. The striped line on the $f$-axis denotes the region where
anomalous diffusion is present. \label{fig:Phasedia}}
\end{figure}

Before turning to the study of finite processivity we review some
of the results which are known for the infinite processivity limit
\cite{Kafri04}. We show later that these results may be reproduced
by studying the spectral properties of the evolution operator of
the probability distribution. It is the spectral method that will
allow us to determine the dynamics most readily when the
processivity is finite. In the homogeneous case ($\alpha,\alpha',
\omega$ and $\omega'$ independent of position), when no disorder
is present the model is described on long times and large
length-scales by a random walker moving along a potential with an
overall tilt between two even sites which is given by
\begin{eqnarray}
\Delta E&=&T\ln\left( \frac{w_a^\leftarrow
w_b^\leftarrow}{w_a^\rightarrow
w_b^\rightarrow}\right) \nonumber \\
&=&T\ln\left( \frac{(\alpha + \omega)(\alpha' e^{\Delta \mu /T} +
\omega')}{(\alpha e^{\Delta \mu /T} + \omega)(\alpha' +
\omega')}\right) +2f \;. \label{eq:dE}
\end{eqnarray}
We refer to such an energy landscape as an effective energy
landscape. It is an alternative description of the dynamics
associated with the rates of Eq. \ref{eq:rates}, which in general
will not satisfy detailed balance since they describe
nonequilibrium processes. Note that when $f=0$ and the chemical
potential difference $\Delta \mu =0$, one has $\Delta E=0$ and no
net motion is generated. Also, when there is directional symmetry
in the transition rates $\alpha=\alpha'$, $\omega=\omega'$
(reflecting directional symmetry in the DNA track) and $f=0$ one
has $\Delta E=0$ even when $\Delta \mu \neq 0$. Absent this
symmetry, chemical energy can be converted to motion and an
effective tilted potential is generated. Similar conditions for
biased motion have been shown to exist for continuum models
\cite{Prost94,Julicher97}. The effect of the externally applied
force is simply to change the overall tilt in the potential. Thus,
for motors moving along a homogeneous (or periodic) polymer, one
expects the velocity to change continuously as $f$ is changed (see
Fig. \ref{fig:Phasedia}).

For motors moving on heterogeneous filaments the situation is
very different. Here the set of parameters
$\{p\}=\{\alpha,\alpha',\omega,\omega',\Delta \varepsilon\}$ is
drawn from a random distribution. Each set of parameters describes
the dynamics on a given type of monomer. Using the results
presented above it is easy to see using equation (\ref{eq:dE})
that the total effective energy change after $m$ monomers is given
by
\begin{equation}
E(m)=\sum_{l=1}^m \Delta E(l)\;. \label{eq:menergy}
\end{equation}
Here, each $\Delta E(m)$ corresponds to an independent set of
values of $\{p\}$ drawn randomly for the m$^{th}$ monomer.
Assuming that $\Delta E(m)$ is drawn from a random distribution
with a finite variance, the effective energy landscape corresponds
to a {\it biased random walk}. Such energy landscapes are
typically referred to as random forcing energy landscapes.

The above scenario applies as long as the chemical potential
difference $\Delta \mu \neq 0$. In the case when $\Delta \mu = 0$
it is easy to see that $E(m)=0$ unless we allow for the energy at
even sites also to vary and take the value $\varepsilon(m)$
\cite{com2}. In this case we obtain
\begin{equation}
E(m)=2fm+\varepsilon(m) \;, \label{eq:renergy}
\end{equation}
corresponding to a {\it random energy} landscape provided
$\varepsilon(m)$ has only short range correlations.

The dynamical behaviors of random walkers in random forcing or
random energy landscapes have been studied in detail
\cite{Derrida83,Bouchaud90}. Using the results of Derrida
\cite{Derrida83} one can calculate the transition points between
the different regimes including the effect of randomness for our
model \cite{Kafri04}. Upon denoting averages over the disorder
rates by an overbar one finds the following regimes.

\noindent {\bf Regime I}: Ordinary biased diffusion occurs when
\begin{equation}
f<-\frac{T}{4} \ln \overline{ \left( \frac{w_a^\leftarrow
w_b^\leftarrow}{w_a^\rightarrow w_b^\rightarrow} \right)^2 }_{f=0}
\;, \label{eq:crf1}
\end{equation}
or
\begin{equation}
f>\frac{T}{4} \ln \overline{ \left( \frac{w_a^\rightarrow
w_b^\rightarrow}{w_a^\leftarrow w_b^\leftarrow} \right)^2 }_{f=0}
\;, \label{eq:crf2}
\end{equation}
where the subscript $f=0$ denotes that $f$ has been set to zero
when evaluating the average. In this regime $ \langle x \rangle
=vt$ and $\langle x^2\rangle - \langle x \rangle^2 =2Dt$ for long
times, where the angular brackets denote an average over different
thermal histories of a particle starting from a particular point.
Another way of stating this result is that for large times the
calculated velocity and diffusion constant, defined by the above
relations, do not depend on the size of the time window, $t_W$,
over which they are evaluated. This results {\it only} holds for
biases satisfying Eqs. (\ref{eq:crf1}) and (\ref{eq:crf2}).

\noindent {\bf Regime II}: The calculated diffusion constant now
depends on the size of the time window, $t_W$, over which it is
evaluated. The velocity does not. Namely, in this region $ \langle
x \rangle =vt$ and $\langle x^2 \rangle - \langle x\rangle^2 \sim
t^{2/\mu}$, where $1<\mu(f)<2$. In the infinite $t_W$ limit the
diffusion constant, $D \equiv \lim_{t_W \to \infty}(\langle
x(t_W)^2 \rangle - \langle x(t_W)\rangle^2)/t_W$, diverges. This
anomaly occurs in the ranges
\begin{equation}
 -\frac{T}{4} \ln \overline{ \left( \frac{w_a^\leftarrow
w_b^\leftarrow}{w_a^\rightarrow w_b^\rightarrow} \right)^2 }_{f=0}
< f \leq -\frac{T}{2} \ln \overline{ \left( \frac{w_a^\leftarrow
w_b^\leftarrow}{w_a^\rightarrow w_b^\rightarrow} \right) }_{f=0}
\;,
\end{equation}
and
\begin{equation}
\frac{T}{2} \ln \overline{ \left( \frac{w_a^\rightarrow
w_b^\rightarrow}{w_a^\leftarrow w_b^\leftarrow} \right) }_{f=0}
\leq f <  \frac{T}{4} \ln \overline{\left( \frac{w_a^\rightarrow
w_b^\rightarrow}{w_a^\leftarrow w_b^\leftarrow} \right)^2 }_{f=0}
\;.
\end{equation}

\noindent {\bf Regime III}: Here {\it both} the velocity and the
diffusion constant are functions of the size of the time window
$t_W$, and $ \langle x \rangle \sim t^\mu$ and $\langle x^2
\rangle - \langle x \rangle^2 \sim t^{2 \mu}$, where $\mu(f)<1$.
In the infinite $t_W$ limit the velocity $v \equiv \lim_{t_W \to
\infty}(\langle x(t_W) - x(0)\rangle)/t_W$ vanishes. The diffusion
constant, $D \equiv \lim_{t_W \to \infty}(\langle x(t_W)^2 \rangle
- \langle x(t_W)\rangle^2)/t_W$, either vanishes or diverges
depending on whether $\mu<1/2$ or $\mu>1/2$. This behavior occurs
when
\begin{equation}
-\frac{T}{2} \ln \overline{\left( \frac{w_a^\leftarrow
w_b^\leftarrow}{w_a^\rightarrow w_b^\rightarrow} \right) }_{f=0}
\leq f \leq \frac{T}{2} \ln \overline{ \left(
\frac{w_a^\rightarrow w_b^\rightarrow}{w_a^\leftarrow
w_b^\leftarrow} \right)}_{f=0} \;.
\end{equation}

\noindent {\bf Sinai diffusion}: Exactly at the stall force $f_s$,
$\langle x \rangle=0$ and $\langle x^2 \rangle \sim
(\ln(t/\tau))^4$, where $\tau$ is the microscopic time needed to
move across one monomer. The ``stall force'' corresponding
to a disordered track is defined by
\begin{equation}
f_s = \frac{T}{2} \overline{ \ln  \left( \frac{w_a^\rightarrow
w_b^\rightarrow}{w_a^\leftarrow w_b^\leftarrow} \right) }_{f=0}
\;.\label{eq:sinailoc}
\end{equation}

\begin{figure}
\includegraphics[width=8cm]{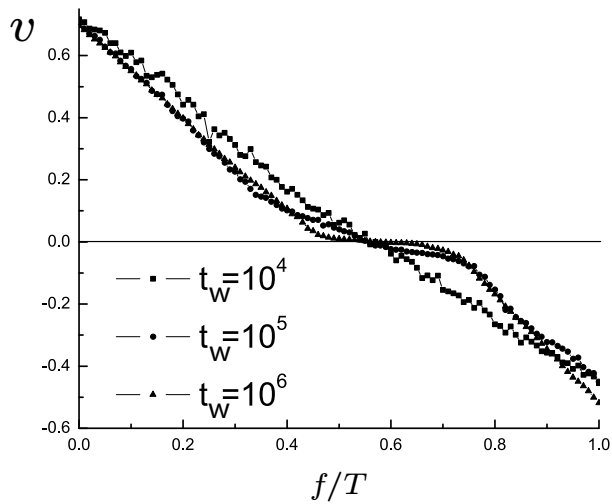} \caption{The velocity as a
function of $f/T$ for different values of $t_W$. Here $\Delta \mu
/T=3$ and parameters were chosen with equal probability to be
either $\{p\}=\{5,1,0.3,1,0\}$ or $\{p\}=\{4,0.1,0.7,1,0\}$ (see
text for notation). The calculated regime of anomalous velocity is
$0.5116<f/T<0.699$. Data incorporates 100 runs (thus averaging
over thermal fluctuations) for a single realization of the
disorder. \label{fig:simu}}
\end{figure}

\begin{figure}
\includegraphics[width=12cm]{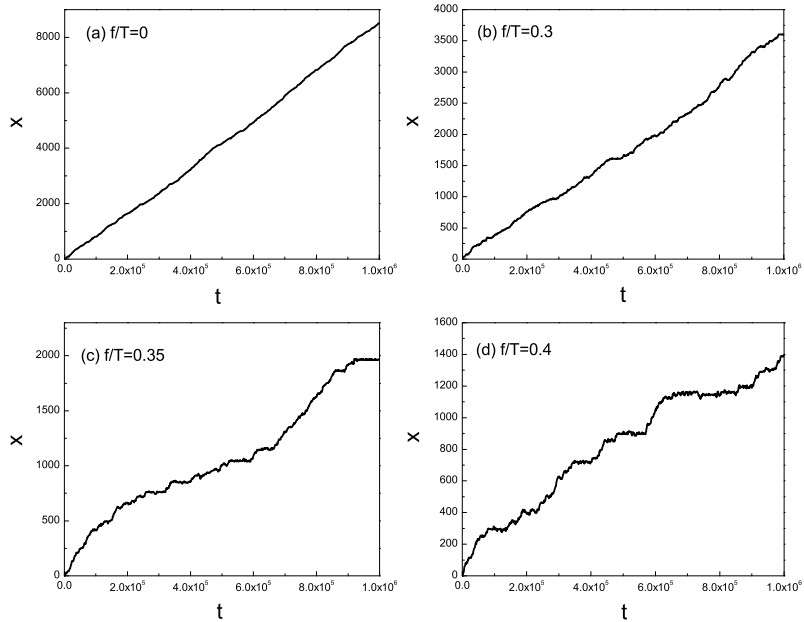} \caption{Typical motor
trajectories shown for the same parameters as Fig. \ref{fig:simu}.
The values of $f/T$ are indicated in the figure. Note that the
plateaus and jumps become more pronounced as $f$ increases toward
the stall force $f_s \simeq 0.56$\label{fig:traces}}
\end{figure}

The resulting behavior as the force is varied is summarized
qualitatively in Fig. \ref{fig:Phasedia} in the limit of an
infinite averaging time window $t_W$. Most notable is the region
of forces (region III) over which the displacement of the motor is
sublinear and the usual long time velocity vanishes. Experiments
are performed with finite $t_W$. The measured velocity then
behaves in region III as $t_W^{\mu-1}$, smoothing the curve shown
in Fig. \ref{fig:Phasedia}, and naturally giving rise to a {\it
convex} shape of the velocity-force curve. This convexity is
demonstrated in Fig. \ref{fig:simu} where the model was simulated
and the velocity measured using different averaging time windows
$t_W$ (see Appendix A for details). As is evident in this figure,
the larger $t_W$, the closer is the velocity-force curve to that
shown in Fig. \ref{fig:Phasedia}. In Fig. \ref{fig:traces} typical
trajectories of the motor on the track are shown for different
values of $f/T$ for a single realization of the disorder. Plateaus
and jumps appear as one moves closer to the region of anomalous
displacement, and the dynamics is controlled by deep minima in the
effective energy landscape with rapid transitions between them.
Such motion is typical of random forcing energy landscapes
\cite{Bouchaud90}.

\section{Infinite Processivity: A Spectral Analysis Approach}
\label{sect:inf-proc-spectr}

The behavior of the infinite processivity limit, as summarized above,
is rather well understood. In this section we show that the dynamics
with infinite processivity may also be deduced by an alternative
approach: the spectral properties of the evolution operator of the
probability density are exploited to deduce the long time, large
length-scale properties of the model. The important features of the
eigenvalues and eigenfunctions which characterize the long time
dynamics are expected to be insensitive to details of the model.  The
method has previously been applied in the study of the physics of
vortex lines in superconductors \cite{HatanoNelson96,HatanoNelson97}
and population dynamics \cite{NelsonShnerb98,Dahmen00}. Earlier
studies considered the spectral properties of a random walker on a
random forcing energy landscape after ``gauging away'' the external
bias, making the evolution operator hermitian
\cite{Bouchaud90}. However, as we will show, a more direct approach
reveals important features (such as complex eigenvalue spectra) which
are not easily observed in the approach of \cite{Bouchaud90}. Although
here the approach is used to reproduce known results, it might prove
useful in the study of more complicated models with an infinite
processivity where an analytic solution is not possible.
%For example, in models where nontrivial correlations of the
%heterogeneity in the rates exist (as can be the case for
%non-coding DNA \cite{Stanley}) and an analytic solution is not
%available.
In addition, the spectral approach to the infinite processivity
model will be very useful as a benchmark once we address the
question of finite processivity.

In what follows, we will often consider a coarse-grained effective
dynamics in which each new lattice site represents a unit cell,
containing one {\it a} and one {\it b} site, of the original
lattice (Eq.~\ref{eq:rates} and Fig.~\ref{fig:motor}).  The
lattice spacing in this new coarse-grained model is $a = 2 a_0$.
Below we first review the spectrum (and the associate dynamics) of
the trivial homogeneous model before turning to the disordered
one.

\subsection{Homogeneous Model}
As discussed above the homogeneous problem is well described on
long time-scales and large length-scales by a random walker
moving along a tilted potential. To see this from the model's spectrum, we first solve for the probability of being in odd sites and
substitute the solution in the equation for the probability of
being in even sites. The resulting coarse-grained equation \cite{Kafri04}, in the
long-time limit, for the probability, $P(x,t)$, of being at site
$x$ at time $t$ is given by
\begin{eqnarray}
(w_b^\rightarrow+w_b^\leftarrow)\partial_t P(x,t)&=&w_a^\rightarrow w_b^\rightarrow P(x-2,t)+w_a^\leftarrow w_b^\leftarrow P(x+2,t) \nonumber \\
&-&(w_a^\rightarrow w_b^\rightarrow +w_a^\leftarrow
w_b^\leftarrow)P(x,t)\;. \label{eq:homo}
\end{eqnarray}
In the continuum limit, which describes the long time, large
length-scale behavior of the model we have
\begin{equation}
\partial_t P(x,t) = D \partial_x^2 P(x,t) -v\partial_xP(x,t) \;,
\end{equation}
with $D=a^2(w_b^\rightarrow w_a^\rightarrow+w_b^\leftarrow
w_a^\leftarrow)/2(w_b^\rightarrow+w_b^\leftarrow)$ and
$v=a(w_b^\rightarrow w_a^\rightarrow-w_b^\leftarrow
w_a^\leftarrow)/(w_b^\rightarrow+w_b^\leftarrow)$. Here, as noted above, $a=2a_0$
is a lattice constant analogous to a unit cell size in conventional
solid state physics. We consider a system of size $L = Na$ and perform
a Laplace transform in time and a Fourier transform in space so
that
\begin{equation}
P(x,t)=\sum_{k} g(k) e^{ikx+\lambda(k)t} \;.
\end{equation}
The coefficient $g(k)$ will depend on the initial conditions. If
periodic boundary condition are imposed on a lattice of size $L$,
the wavevectors $k$ specifying the (delocalized) eigenfunctions
$\Psi_k(x) \sim e^{ikx}$ are quantized, $k_n=2\pi n /L$, $n=0,
\pm1, \ldots$. In general, since the evolution operator is
non-hermitian eigenvalues may be complex. The complex eigenvalues
must come in complex conjugate pairs to ensure a real probability
density $P(x,t)$ \cite{HatanoNelson97}. The eigenvalue spectrum is
given by
\begin{equation}
\lambda(k) = -D k^2-i v k \;, \label{eq:hspec}
\end{equation}
corresponding to diffusion with drift.  The dependence of the real
part of the spectrum on $k$ describes diffusion, while the
dependence of the imaginary part describes drift with a constant
velocity. This identification is possible when periodic boundary
conditions allow a current to flow in the system even in the limit
of $t \to \infty$.

\subsection{Heterogeneous Model}
For the heterogeneous model the situation is more subtle, because
the Fourier transform cannot be used to diagonalize the evolution
operator. Nevertheless, we will show that a numerical analysis of
the eigenvalue spectrum can be used to deduce the dynamical
properties of the model. The role of the wave numbers in the
homogeneous model is taken by {\it winding numbers}
\cite{ShnerbNelson98}.

The winding number, defined only for complex eigenfunctions, is
given by the number of times the eigenfunction spirals around the
origin in the complex plane as it traverses the whole lattice
with periodic boundary conditions. For homogeneous systems and
eigenfunctions which behave like $e^{ik_nx}$ we have $k_n=2\pi
n/Na$, implying a winding number $n$. For disordered systems, it can be shown
\cite{ShnerbNelson98} that the winding number increases linearly
with the {\it eigenvalue index} $n$ which orders the
eigenfunctions from the lowest value of $|\lambda(n)|$ to the
highest. That is, the winding number associated with an eigenvalue
of magnitude $|\lambda(n)|$ is $\pm n$. Moreover, if the
eigenvalue $\lambda(n)$ is in the upper part of the complex
$\lambda$ plane its winding number is $n$. Its complex conjugate
pair $\lambda^*(n)$ will then have a winding number $-n$. For real
eigenvalues the eigenfunctions can be chosen to be real and the
winding number is not defined. In this case, the eigenfunctions
can be classified by the number of zeros, as in one dimensional
quantum mechanics \cite{Landau}.

The eigenvalue spectrum of the heterogeneous model with periodic
boundary conditions can be used to encapsulate the long time
dynamical properties of the model. Specifically, the dependence of
the real and imaginary parts of the eigenvalues on the {\it
winding number} signals the properties of the mean square
displacement and drift of the motor molecule within our model.

To this end we consider the model of Sec. II with, as before, two
types of monomers drawn at random. The set of parameters
representing each type of monomer
$\{p\}=\{\alpha,\alpha',\omega,\omega',\Delta \varepsilon \}$ is
chosen with probability $1/2$ to be
$\{p_1\}=\{\alpha_1,\alpha_1',\omega_1,\omega_1',\Delta
\varepsilon_1\}$ and with probability $1/2$ to be
$\{p_2\}=\{\alpha_2,\alpha_2',\omega_2,\omega_2',\Delta
\varepsilon_2\}$. The chemical potential difference $\Delta \mu$
is assumed to be the same for both types of monomers. Allowing
$\Delta \mu$ to depend on the type of monomer does not alter the
qualitative long time behavior of the model.

We calculated the eigenvalue spectrum of the evolution operator
numerically by diagonalizing a matrix with a specific realization
of the disorder (see Appendix B for more details). From the exact
solution of the model we expect three different regimes. Below a
typical spectrum from each of the regimes is shown and examined in
detail. Throughout we set $T=1$ and use the parameters
$\{p_1\}=\{\alpha_1,\alpha_1',\omega_1,\omega_1',\Delta
\varepsilon_1 \}=\{6,1,1,6,0\}$ and
$\{p_2\}=\{\alpha_2,\alpha_2',\omega_2,\omega_2',\Delta
\varepsilon_2\}=\{1.2,1,1,1.2,0 \}$ (again with equal probability)
and $e^{\Delta \mu /T}=10$. The results are unchanged for similar
sets of parameters. For these values of the parameters the exact
locations of the transitions between the different regimes can be
readily calculated using the results summarized in Sec. II.

\begin{figure}
\includegraphics[width=8cm]{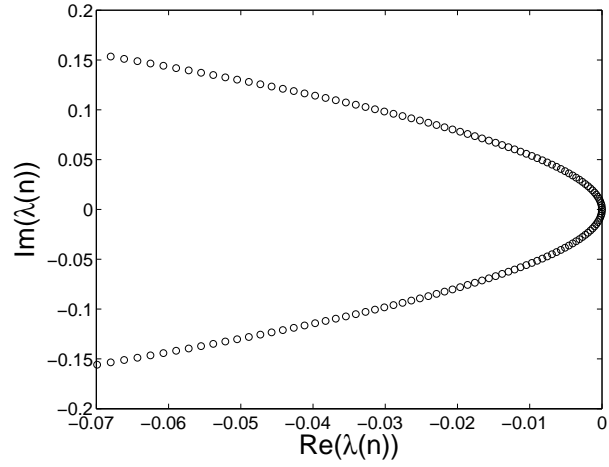} \caption{The eigenvalues
obtained for $f=0$. Here the system size is $N=4500$. Shown are the
$140$ eigenvalues with the lowest value of $|\lambda|$.   In the region
near the origin we expect that ${\rm Im} (\lambda) \propto \pm
\sqrt{ |{\rm Re} (\lambda)| }$.}
\label{fig:Ef0}
\end{figure}

\noindent {\bf Regime I:} For $f<0.2256$ and $f>0.518$ the motion
is biased diffusion. A typical spectrum of eigenvalues in the
complex plane (for $f=0$) is shown in Fig. \ref{fig:Ef0}. We show
only the eigenvalues with the smallest $| \lambda|$, because those
with larger values are non-universal and are highly dependent on
the details of the model used to describe the motion of the motor.
The spectrum has two branches in the complex $\lambda$ plane.
Also, it has no purely real parts (with the exception of
$\lambda=0$).

For usual diffusion with uniform drift in a homogeneous system, as
discussed above, $\lambda(k)=-Dk^2-ivk$.  The imaginary part of
the spectrum corresponds to drift, the real part to diffusion. A
similar behavior might be expected in our heterogeneous model for
$\lambda(k)$ where $k$ is related to the winding number as
$k=k_n=2\pi n/Na$.  This can be seen in a plot of the real and
imaginary parts of the spectrum as a function of the eigenvalue
index as defined above. Since in this regime the dynamics is
biased diffusion we expect ${\rm Im}(\lambda(n)) \propto n$, while
${\rm Re}(\lambda(n)) \propto n^2$.

\begin{figure}
\includegraphics[width=8cm]{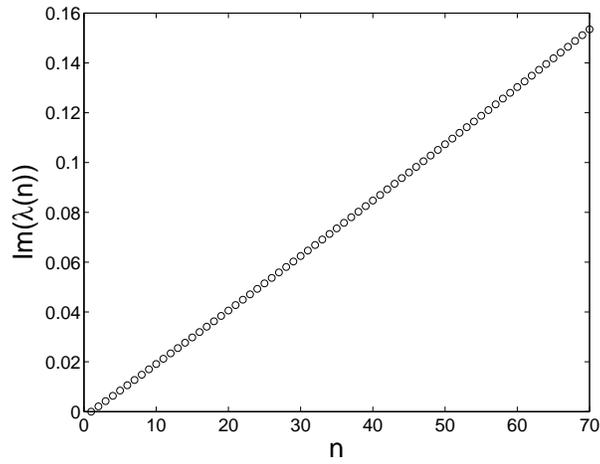} \caption{The imaginary part
of the eigenvalue, ${\rm Im}(\lambda(n))$ for $f=0$ as a function
of $n$. Here the system size is $N=4500$. Shown are the $140$
eigenvalues with the lowest value of $|\lambda|$.}
\label{fig:Ef0v}
\end{figure}

\begin{figure}
\includegraphics[width=8cm]{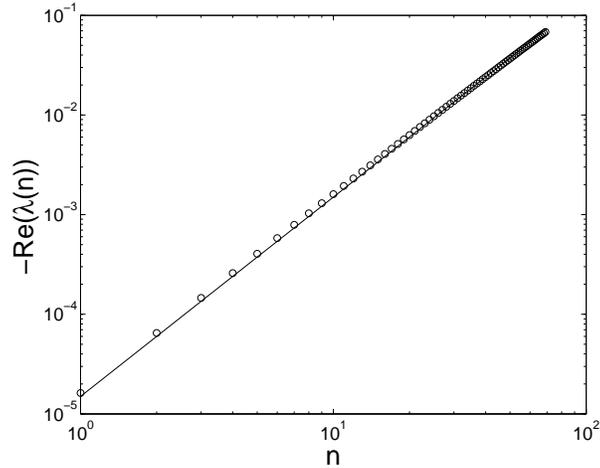} \caption{The real part
of the eigenvalue, $-{\rm Re}(\lambda)$ for $f=0$ as a function of
the winding number $n$. Here the system size is $N=4500$. Shown
are the $140$ eigenvalues with the lowest value of $|\lambda|$.
The solid line is the function $An^2$ with $A$ a constant.}
\label{fig:Ef0D}
\end{figure}

To test this hypothesis, we order the eigenvalues according to
their magnitude and plot ${\rm Im}(\lambda(n))$ and ${\rm
Re}(\lambda(n))$. Fig. \ref{fig:Ef0v} shows ${\rm Im}(\lambda(n))$
for small $n$. One can see that indeed the slope in linear. Fig.
\ref{fig:Ef0D} shows $-{\rm Re}(\lambda(n))$ for small $n$ on a
log-log scale along with a line $-{\rm Re}(\lambda(n))=An^2$ with
$A$ some constant. At small $n$ indeed $Re(\lambda(n)) \propto
n^2$ for more than a decade.

\begin{figure}
\includegraphics[width=8cm]{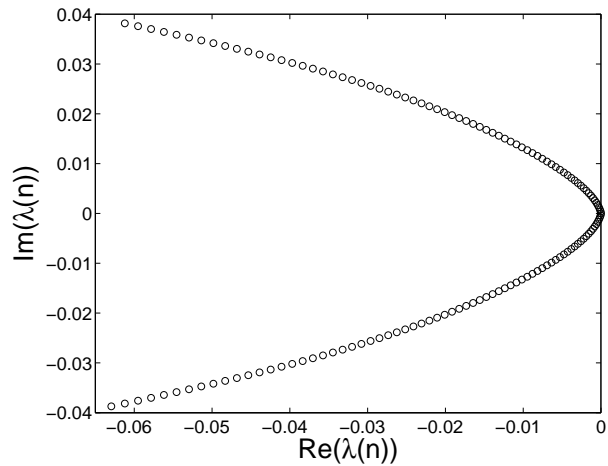} \caption{The eigenvalues
obtained for $f=0.25$. Here the system size is $N=4500$. Shown are
the $140$ eigenvalues with the lowest value of $|\lambda|$.  Near
the origin we expect that ${\rm Im} (\lambda) \propto \pm |{\rm
Re} (\lambda)|^{1/\mu}$, with $1<\mu<2$.} \label{fig:Ef027}
\end{figure}

\begin{figure}
\includegraphics[width=8cm]{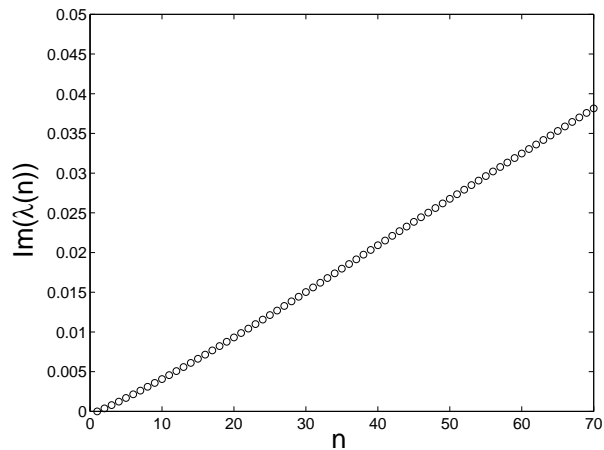} \caption{The imaginary part
of the eigenvalue for $f=0.25$ as a function of the winding number
$n$. Here the system size is $N=4500$. Shown are the $140$
eigenvalues with the lowest value of $|\lambda|$.}
\label{fig:Ef027v}
\end{figure}

\begin{figure}
\includegraphics[width=8cm]{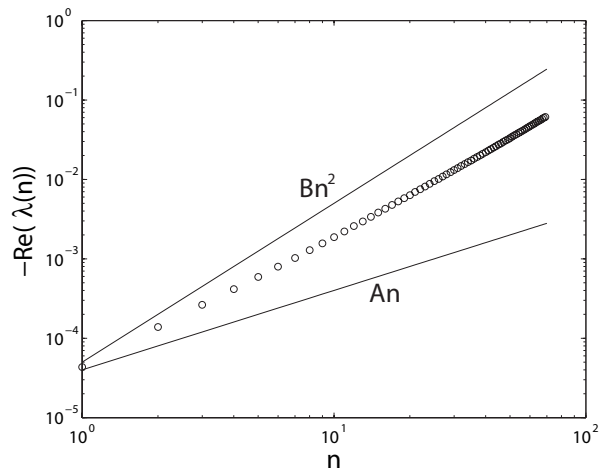} \caption{The real part
of the eigenvalue, $-Re(\lambda)$ for $f=0.25$ as a function of
$n$. Here the system size is $N=4500$. Shown are the $140$
eigenvalues with the lowest value of $|\lambda|$. The function
$An$ and $Bn^2$ with $A$ and $B$ constants are plotted for
reference. From a fit to the slope of our numerical data we find
$\mu=1.915 \pm 0.004$.} \label{fig:Ef027D}
\end{figure}

\noindent {\bf Regime II:} Next we turn to look at the region
where the diffusion constant depends on the size of the time
averaging window used to evaluate it, while the velocity does not.
For the parameters used in the numerics this occurs for force
windows given by $0.2256<f<0.2882$ and $0.4555<f<0.518$. A typical
spectrum of eigenvalues in the complex plane is shown in Fig.
\ref{fig:Ef027}. Again we concentrate on the small $| \lambda |$
part of the spectrum and examine ${\rm Im}(\lambda(n))$ and
$Re(\lambda(n))$. Fig.  \ref{fig:Ef027v} presents a plot of ${\rm
Im}(\lambda(n))$ for small $n$. The expected dependence on $n$ is
linear. A careful analysis of the curve shows that ${\rm
Im}(\lambda(n))$ can be fit well to $An+Bn^3$ for small $n$ (here
we have used that fact that ${\rm Im}(\lambda(n))$ is expected to
be an odd function of $n$) . The coefficient $B$ decreases as the
size of the system studied is increased. Such a correction is
expected if the time to reach the asymptotic behavior $\langle x
\rangle \sim v t$ is so large that it is comparable to the
relaxation time of the system. For larger systems the correction
will become less important and the asymptotic behavior will be
observed more easily. Fig. \ref{fig:Ef027D} shows
$-Re(\lambda(n))$ for small $n$ on a log-log scale along with the
lines $An$ and $Bn^2$ with $A$ and $B$ some constants.  At small
$n$, $Re(\lambda(n))$ indeed does not behave as $n^2$. It shows a
smaller slope in accordance with a behavior consistent $n^\mu$
with $1<\mu<2$. In fact, from naive dimensional considerations
($\lambda$ behaves as $1/t$ while $x \sim n$) this is the behavior
expected when $\langle x^2 \rangle-\langle x \rangle^2 \sim
t^{2/\mu}$.

\begin{figure}
\includegraphics[width=8cm]{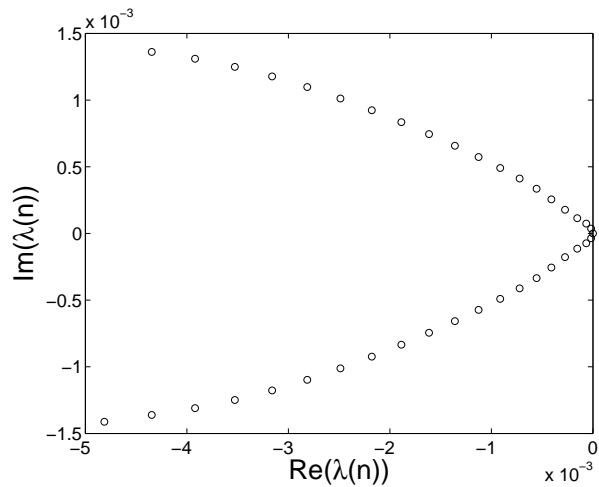} \caption{The eigenvalue spectrum obtained for
$f=0.31$, where both diffusion and drift are anomalous. Here the
system size is $N=4500$. Shown are the $40$ eigenvalues with the
lowest value of $|\lambda|$. Near the origin, we expect ${\rm
Im}(\lambda) \propto \pm |{\rm Re}(\lambda)|$.} \label{fig:Ef0305}
\end{figure}

\begin{figure}
\includegraphics[width=8cm]{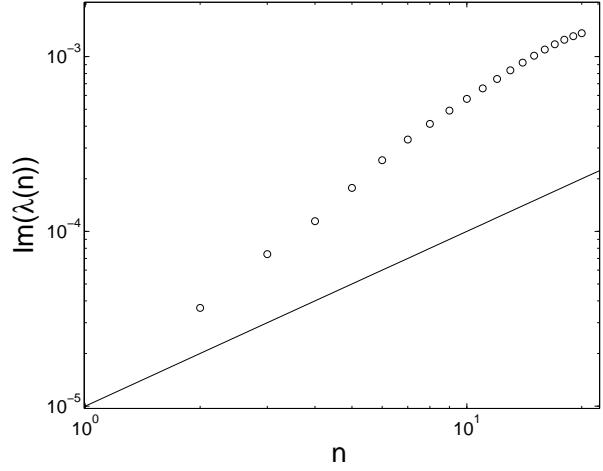} \caption{The imaginary part
of the eigenvalue for $f=0.31$ as a function of the winding number
$n$. Here the system size is $N=4500$. Shown are the $40$
eigenvalues with the lowest value of $|\lambda|$. A fit to ${\rm
Im}(\lambda) \propto -|n|^{1/\mu}$, over only one decade, gives
$\mu=0.6 \pm 0.1$. The function $An$ with $A$ a constant is
plotted for reference.} \label{fig:Ef0305v}
\end{figure}

\noindent {\bf Regime III:} Finally, we turn to look at the region
where the velocity is effectively zero as $t_W \to \infty$ and the
diffusion is again anomalous. For the parameters used in our
numerical analysis this regime occurs for $0.2882<f<0.518$. A
spectrum of eigenvalues in the complex plane is shown in Fig.
\ref{fig:Ef0305}. Again we concentrate on the small $| \lambda |$
part of the spectrum. To get a clear picture one needs to plot
both ${\rm Im}(\lambda(n))$ and ${\rm Re}(\lambda(n))$. The
positive part ${\rm Im}(\lambda(n))$ is plotted on a log-log plot
in Fig.~\ref{fig:Ef0305v}. Here the displacement of the particle
is expected to behave as $t^{\mu}$ with $\mu<1$. This leads to an
expected dependence ${\rm Im}(\lambda(n)) \propto n^{1/\mu}$.  A
reference line $An$ with $A$ a constant is plotted for comparison.
Clearly, ${\rm Im}(\lambda(n))$ behaves as expected.

\begin{figure}
\includegraphics[width=8cm]{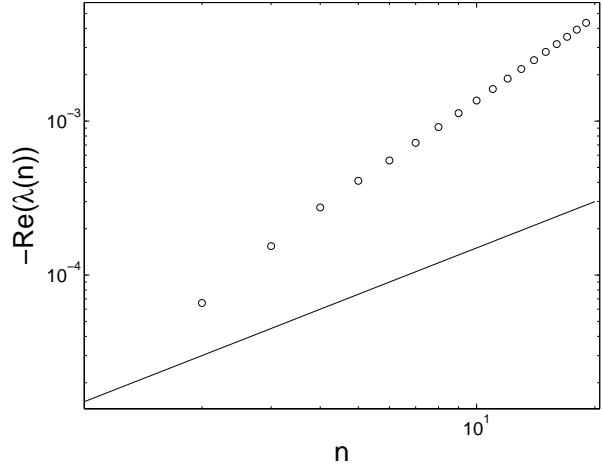} \caption{The real part
of the eigenvalue, $-Re(\lambda)$ for $f=0.31$ as a function of
$n$. Here the system size is $N=4500$. Shown are the $40$
eigenvalues with the lowest value of $|\lambda|$. A fit to ${\rm
Re}(\lambda) \propto n^{1/\mu}$, over only one decade, gives
$\mu=0.6 \pm 0.1$. The function $An$ with $A$ a constant is
plotted for reference.} \label{fig:Ef0305D}
\end{figure}

Fig. \ref{fig:Ef0305D} shows $-{\rm Re}(\lambda(n))$ for small $n$
on a log-log scale along with a line $An^2$ with $A$ some
constant. As can be seen at small $n$ indeed ${\rm
Re}(\lambda(n))$ does not behave as $n^2$. Here a similar argument
as before leads one to expect $-{\rm Re}(\lambda(n)) \propto n^{1/\mu}$
for small $n$, consistent with the numerics.

We comment that the analysis is only possible not too far from the
transition point into regime III. Deep inside the region with
anomalous velocity, the small $|\lambda|$ spectrum becomes very
noisy due to finite size effects. Analysis of the spectrum then
becomes difficult due to these strong sample-to-sample
fluctuations. Hence we have not attempted an analysis in the Sinai
diffusion regime, where the energy landscape is nearly horizontal.

\section{Finite Processivity}
\subsection{Homogeneous Model}

As discussed above, the homogeneous model is well described on
long time scales and large length scales by a random walker moving
along a tilted potential. The unbinding of motors from the track
adds a non-conserving term to the equation for the probability
density. In a continuum description, one now has
\begin{equation}
\partial_t P(x,t)=D\partial_x^2 P(x,t)-v\partial_x P(x,t) - w_{\rm
off}P(x,t) \;. \label{eq:homofall}
\end{equation}
In terms of the microscopic model as before one has
$D=a^2(w_b^\rightarrow w_a^\rightarrow+w_b^\leftarrow
w_a^\leftarrow)/2(w_b^\rightarrow+w_b^\leftarrow)$ and
$v=a(w_b^\rightarrow w_a^\rightarrow-w_b^\leftarrow
w_a^\leftarrow)/(w_b^\rightarrow+w_b^\leftarrow)$. In addition,
here $w_{\rm off}=(w_{\rm
off}^b(w_a^\rightarrow+w_a^\leftarrow)+w_{\rm
off}^a(w_b^\rightarrow+w_b^\leftarrow))/(w_b^\rightarrow+w_b^\leftarrow)$.

Again, we consider the model with periodic boundary conditions and
perform a Laplace transform in time and a Fourier transform in
space. The eigenvalue spectrum is given by $\lambda(k) = -D k^2-i
v k -w_{\rm off}$. When $w_{\rm off}=0$ the motion is diffusion
with drift. However, when $w_{\rm off}>0$ the probability density
decays exponentially to the empty track state $P(x,t)=0$ with a
typical time scale $1/w_{\rm off}$. Note, however, that the
eigenfunctions are identical to the infinite processivity case. In
fact, the probability density of the motor can be written as
\begin{equation}
P(x,t)=e^{-w_{\rm off}t}P_{w_{\rm off}=0}(x,t) \;.
\end{equation}
Up to a rescaling of the probability density the time evolution is
unaltered. That is, the motors which remain on the track are {\it
unaffected} by a constant hopping off rate.

\subsection{Heterogeneous Model}

If the hopping off rates are uniform along the track, but the
remaining parameters are chosen randomly, $w^a_{\rm
off}$ and $w^b_{\rm off}$ are the same for all types of monomers,
and it is straightforward to see that the effect is the same as
for a homogeneous system: the dynamics of the motors which remain
on the track are unaffected by the non-conservation.

We now show that, when randomness in the rates for hopping off the
track is introduced into the system, the dynamics of the motors
which remain on the track can be altered in a dramatic way. The
eigenfunctions of the evolution operator are affected by the
heterogeneous hopping off rates as well as by the sequence
heterogeneity as it affects the local diffusion constant and drift
velocity. As we shall see, disorder in the hopping off rates has a
profound effect on the behavior of motors which remain on the
track. We stress that, with applications to molecular motors in
mind, the disorder in the detachment rates we consider is {\it
correlated} with the local hopping rates.

As for the conserving heterogeneous model with its random force
landscape, we study the dynamics by considering the spectral
properties of the evolution operator of the model. The effect of
detachment is included by setting the rates $w_{\rm off}^a$ and
$w_{\rm off}^b$, corresponding to leaving the track at even and
odd sites respectively, to be non-zero and site dependent. We find
two types of behavior which depend on the drift properties of the
motor for the corresponding conserving model. When the drift is
linear in time the last motors to be left on the track localize
only beyond a critical disorder strength in the hopping off rates.
Similar behavior arises in the physics of vortices
\cite{HatanoNelson96,HatanoNelson97} and in population dynamics
\cite{Dahmen00}. The behavior when the displacement of the motor
for the conserving model is sublinear (i.e. when it is not
possible to define a drift velocity) is different. Now, the last
motors to remain on the track are {\it always} localized for any
strength of the disorder.

\subsubsection{Ballistic displacement}

First, we consider the regime where the displacement of the motor
 in the conserving model is linear as a function of time ($\mu>1$).
We will show that when the disorder in the rates of leaving the
track is strong enough the eigenvalues with the smallest
$|\lambda|$ become purely real. This is a regime of localized
eigenfunctions and corresponds to motors with zero drift velocity.
This feature is present over the whole region where $\mu>1$.

\begin{figure}
\includegraphics[width=8cm]{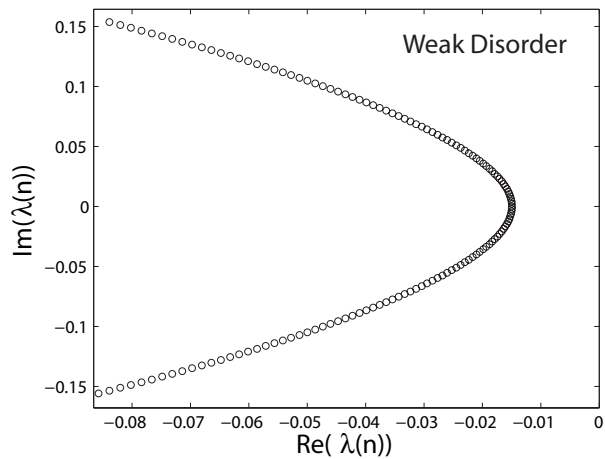} \caption{The eigenvalue spectrum for $f=0$ and $\mu>2$ with rates for hopping off the
track as specified in the text. Except for the largest eigenvalue,
all eigenvalues have an imaginary part. All eigenfunctions,
moreover describe extended states with a well-defined winding
number. Shown are the $140$ eigenvalues with the lowest value of
$|\lambda|$ and the system size is
$N=4500$.\label{fig:spectrumf0fall}}
\end{figure}

\begin{figure}
\includegraphics[width=8cm]{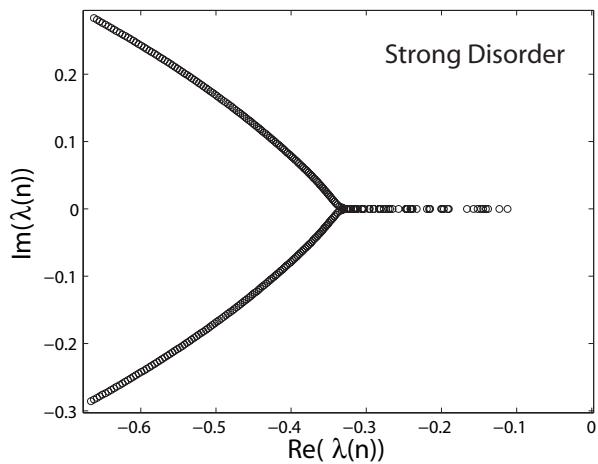} \caption{The eigenvalue spectrum for $f=0$ with rates for hopping off the
track as specified in the text. As can be seen the eigenvalues
with the lowest value of $|\lambda|$ become localized. Shown are
the $300$ eigenvalues with the lowest value of $|\lambda|$ and the
system size is $N=4500$.\label{fig:spectrumf0loc}}
\end{figure}

To explore this behavior, we have studied the eigenvalue spectrum
of the model numerically for systems of size $4500$. In Fig.
\ref{fig:spectrumf0fall} results are shown for the case $f=0$
($\mu>2$). However, the localization effects which are of interest
to us here remain unchanged for all values of $f$ such that
$\mu>1$. The parameters used are the same as used previously in
the paper augmented by $w_{\rm off}^a=0.01$ and $w_{\rm
off}^b=0.01$ ($w_{\rm off}^a=0.02$ and $w_{\rm off}^b=0.02$) for
the parameter set $\lbrace p_1 \rbrace$ ($\lbrace p_2 \rbrace$).
As can be seen, except for the lowest $| \lambda |$ eigenvalue,
all eigenvalues have an imaginary part. Thus, the motors are
biased to move across the lattice in a positive direction. In Fig.
\ref{fig:spectrumf0loc} we show the spectrum with $w_{\rm
off}^a=0.01$ and $w_{\rm off}^b=0.01$ ($w_{\rm off}^a=0.08$ and
$w_{\rm off}^b=0.08$) for the set $\lbrace p_1 \rbrace$ ($\lbrace
p_2 \rbrace$), which corresponds to a larger disorder strength.
Here eigenvalues beyond a certain mobility edge become real,
implying that the long lasting motors are described by localized
states \cite{HatanoNelson97,NelsonShnerb98}. Note, that the
disorder in the hopping off rates is correlated with the disorder
in the hopping rates themselves, as required if both are due to
the same underlying heterogeneous polynucleotide sequence.

The transition between moving and localized motors can be
understood, in the case of ballistic displacement, by adapting the
ideas of Ref. \cite{Grass82}. We consider a simplified model for
the motors. The energy landscape consists of two slopes $s_1$ and
$s_2$ (corresponding to two types of monomers). The velocity on a
track with a slope $s_1$ ($s_2$) is denoted be $v_1$ ($v_2$). The
rate for leaving the track on a region with a slope $s_1$ is
assumed to be $w$, while the rate for leaving the track on a
region with slope $s_2$ is $0$. Note that as before the local
detachment rates are correlated with the local biases.

Next, compare two ``survival'' strategies for the motor in the long
time limit. In the first it
spreads across the lattice probing regions with slope $s_1$ and
$s_2$. In the other it remains in regions with slope $s_2$ where
detachment rates are zero. The probability of staying in such
regions can be estimated by solving for a random walker moving on
a tilted potential of some size with absorbing boundary conditions
at both ends. One finds \cite{Grass82} that this probability
behaves as
\begin{equation}
P_{\rm stay}(t) \sim e^{-v_2^2t/4D } \;, \label{eq:Pstay}
\end{equation}
where $D$ is the diffusion coefficient on a track with slope
$s_2$. The
probability of survival of a particle moving across the lattice
can also be evaluated. One expects
\begin{equation}
P_{\rm move}(t) \sim e^{-c w t} \;, \label{eq:Pmove}
\end{equation}
where $c$ is a constant which will depend on the disorder averaged
velocity.  For large enough $w$, the particles that choose the
localized strategy clearly will have a better chance of survival.
One therefore expects that as the disorder in the hopping off
rates is increased a transition from localized states to
delocalized states will occur at the high end of the spectrum
(corresponding to small $k$ and small winding numbers). Note that
the argument would not be altered if the rate for hopping off from
negative slope regions were non-zero, although, the transition
point between the two regimes would be shifted.

Finally we comment that this argument could also be used for a
model with a random energy landscape (see Eq. \ref{eq:renergy}).
This case is very similar to problems which have been studied in
the context of the vortex physics
\cite{HatanoNelson96,HatanoNelson97} and population dynamics
\cite{NelsonShnerb98}. For completeness, Appendix C presents
eigenvalue spectra, for weak and strong disorder, in that case.

\subsubsection{Sublinear displacement}

\begin{figure}
\includegraphics[width=8cm]{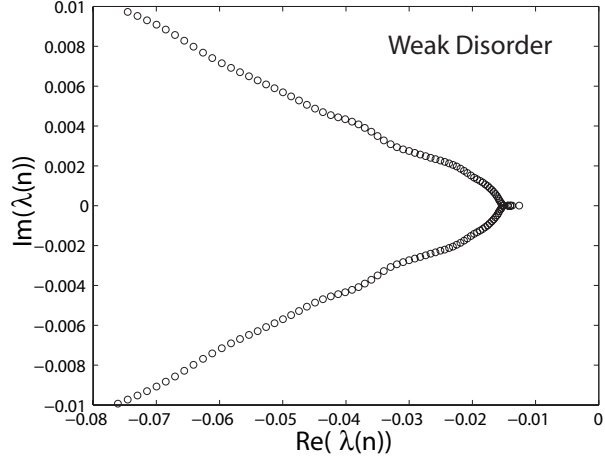} \caption{The eigenvalue spectrum for $f=0.31$ with rates for hopping off the
track as specified in the text. The small band of real eigenvalues
at the top of the spectrum characterizes localized states. Shown
are the $140$ eigenvalues with the lowest value of $|\lambda|$, and
the system size is again $N=4500$.\label{fig:spectrumf031}}
\end{figure}

\begin{figure}
\includegraphics[width=8cm]{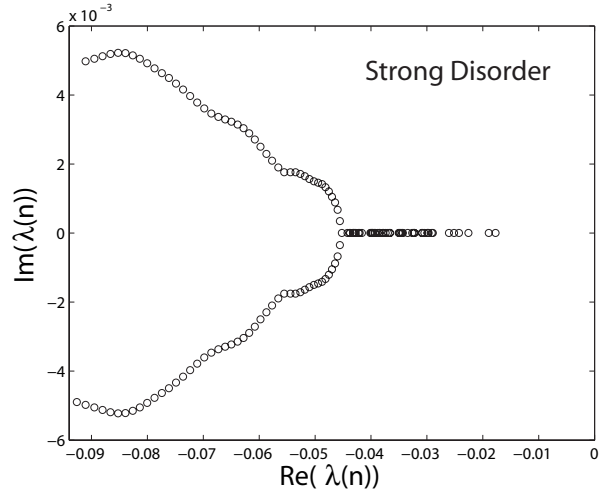} \caption{The eigenvalue spectrum for $f=0.31$ with rates for hopping off the
track as specified in the text. The spectrum is qualitatively
similar to that of Fig. \ref{fig:spectrumf031} with weak disorder.
The band of real eigenvalues at the top of the spectrum
characterizes localized states and now contains, as expected, more
states than Fig. \ref{fig:spectrumf031}. Shown are the
$140$ eigenvalues with the lowest value of $|\lambda|$, and the
system size is again $N=4500$.\label{fig:spectrumf031B}}
\end{figure}

A typical spectrum in the regime of sublinear displacement,
$\mu<1$, is shown in Fig. \ref{fig:spectrumf031}. Note the small
band of localized states with real eigenvalues near the top of the
spectrum (small $|\lambda|$). Again the parameters used are the
same as used previously with the falling from the track rates
$w_{\rm off}^a=0.01$ and $w_{\rm off}^b=0.01$ ($w_{\rm
off}^a=0.02$ and $w_{\rm off}^b=0.02$) for the set $\lbrace p_1
\rbrace$ ($\lbrace p_2 \rbrace$). Note, that even for these small
values of the falling off rates localized states appear. This
effect persists even for falling off rates smaller by more than an
order of magnitude and for various values of $f$ in this regime as
long as large enough systems were studied. For comparison Fig.
\ref{fig:spectrumf031B} shows the spectrum in the case of strong
disorder. Here $w_{\rm off}^a=0.01$ and $w_{\rm off}^b=0.01$
($w_{\rm off}^a=0.06$ and $w_{\rm off}^b=0.06$) for the parameter
set $\lbrace p_1 \rbrace$ ($\lbrace p_2 \rbrace$). As always in
this paper, the two parameter sets are assumed t ooccur with equal
probability. As can be seen from the figure, while details of the
spectrum are different from the weak disorder case shown in Fig.
\ref{fig:spectrumf031} they are qualitatively the same.

\subsubsection{Implications for the probability of finding a motor
on the track as a function of time}

Next, we discuss the effect of the localized states occurring near
the top (low $|\lambda|$) of the spectrum on the probability of
finding a motor on the track as a function of time. It may be
possible to reveal these localized states by looking at the
density decay as a function of time of a dilute concentration of
fluorescently labeled motors placed on a track. The probability of
finding a motor on the track as a function of time, $P_s(t)$,
could then be monitored by looking at the decay of the fluorescent
signal as a function of time (possibly averaging over several
experiments).

The implication of localized states for the probability $P_s(t)$
has already been considered in the context of random walkers
subject to the influence of random traps \cite{Grass82}. The
arguments are unaltered for the case of a random forcing energy
landscape, and here we consider a simple version. A more detailed
proof can be carried out along the lines of Ref.
\cite{NelsonShnerb98}. We consider a system where the hopping and
detachment rates can only assume two values. We are interested in
the behavior in the long-time limit. When localized states exist
the eigenfunctions are strongly peaked in regions where the
probability of detachment from the track is small. The probability
for such a region of length $l$ to occur behaves as $e^{-\gamma
l}$ where $\gamma$ is a constant. On the other hand the time to
leave the region behaves as $e^{-v_2^2 t/4D + \chi D t/l^2}$,
where $\chi$ is a constant of order unity and we have included
higher order corrections to Eq. \ref{eq:Pstay} associated with the
finite size $l$ of the region. Note that we have used the fact
that the detachment rate is correlated with the local hopping rate
so that the behavior inside the region is diffusive irrespective
of the value of $\mu$. Summing over the contribution from regions
of different lengths one obtains
\begin{equation}
P_s(t) \sim e^{-v_2^2t/4D  - (t/t_0)^{1/3}} \;. \label{eq:str}
\end{equation}
Here $t_0$ depends on the constants $\gamma$ and $\chi$ and $D$
the diffusion coefficient. As expected the leading behavior is
still exponential. However, a signature of the localized states
appears in the correction which has a stretched exponential form.

In contrast, when no localized states exist the behavior is
dominated by the detachment rates. In that case no diffusive
corrections to the detachment rate from a region of size $l$ are
present. The stretched exponential correction to the decay of
$P_s(t)$ displayed in Eq. (\ref{eq:str}) will be absent.

In summary, in the region of sublinear displacement one expects a
stretched exponential behavior to always be present at long times
when the probability of finding a motor on the track is monitored.
In the ballistic displacement regime the correction will be
present only for strong enough disorder in the detachment rates.

\section{Toy Model}

The understanding of the dynamics of random walkers (with infinite
processivity) on random forcing energy landscapes has been
enhanced by a simple toy model introduced by Bouchaud et. al.
\cite{Bouchaud90}. The model builds on the fact that at finite
tilt the sojourn time $\tau$ at any site is found to have a broad
distribution $\Psi(\tau) \sim \tau^{-(1+\mu)}$ for large $\tau$.
This behavior suggests that the dynamics could be mimicked by a
{\it directed walk} between traps with a broad release time
distribution. Specifically, the model consists of a particle
moving unidirectionally on a lattice with hopping rates $W_k$
between site $k-1$ and $k$ which are drawn from a probability
distribution which satisfies $\Psi(W) \approx \overline{A}
W^{\mu-1}$ as $W \to 0$. Although the equivalence to the original
model cannot be justified rigorously (since back stepping is
ignored), the two models are known to exhibit the same long time
behavior.

Here we analyze the model from a somewhat different perspective
and show that indeed it yields for the infinite processivity limit
the types of spectra described in Section  III. Moreover, we study
the finite processivity model in the limit of {\it weak disorder}
in the rates for leaving the track. Here the rate for leaving the
track at site $k$, $w_k$, is assumed to be uncorrelated with the
hopping rate $W_k$.  If one views the directed model as a
coarse-grained version of the original undirected model, then one
expects that the hopping rates $W_k$ are determined by the depth
of the traps in the energy landscape.  In contrast, since the
particles spend most of their time near the bottom of the traps,
the coarse-grained off rates in the directed model will depend
primarily on the local off rates near the trap bottoms in the
original model, suggesting that the $w_k$ and $W_k$ should indeed
be uncorrelated.  With this simplification, we show that similar
to the more complex models discussed earlier in this paper, {\it
any strength} of disorder in the rates for leaving the lattice
causes the eigenfunctions describing the long-time behavior of the
model to be localized for $\mu<1$. In contrast for $\mu>1$, weak
disorder leaves the eigenfunctions delocalized.

It is convenient to first consider a more general model with
non-zero rates for leaving the track. We derive an equation for
the eigenvalue spectrum of the model and then analyze the spectra
separately in the infinite and finite processivity limits.

The time evolution of the model is described by the set of master
equations
\begin{equation}
\frac{dP_k(t)}{dt}=W_kP_{k-1}-W_{k+1}P_k-w_{k+1}P_k \;.
\end{equation}
After a Laplace transform, the corresponding equations for the
eigenfunctions is then
\begin{equation}
(\lambda+W_{k+1}+w_{k+1})P_k(\lambda)=W_kP_{k-1}(\lambda) \;,
\end{equation}
which yields a recursion relation for the $P_k(\lambda)$ which can
be readily solved. The possible eigenvalues $\lambda$ are then
determined by imposing periodic boundary conditions, yielding
\begin{eqnarray}
1&=&\prod_{i=1}^{N} \frac{W_i}{(\lambda+W_{i}+w_{i})} \nonumber \\
&\sim&  \exp(N \overline{ \left( \ln \frac{W}{\lambda+W+w}
\right)} ) \label{eq:toybc}
\end{eqnarray}
where the last line holds in the limit when the system size $N$ is
taken to infinity and the overline denotes as before average over
disorder. Care must be taken when $\lambda$ is real. In this case
the recursion relation might allow for some $\lambda+W_{k+1}+w_k$
to be zero. In that case $P_k(\lambda)$ can assume any value (to
be set by normalization), while $P_{k-1}(\lambda)=0$. Such
eigenfunction will be referred to as ``localized'', with a
detailed justification given below. When $\lambda$ is complex, we
refer to the eigenfunctions as delocalized.

According to Eq. (\ref{eq:toybc}), we need to calculate
\begin{equation}
R(\lambda)=\overline{ \left( \ln \frac{W}{\lambda+W+w}  \right)}.
\label{eq:defnR}
\end{equation}
Since we are considering the long time behavior and small rates
for leaving the track we consider the limit $|\lambda+w| \ll 1$.
It will be useful to hold $w$ fixed for now and average over it
later. We denote the function obtained by averaging only over $W$
in Eq.~(\ref{eq:defnR}) by $Q(\lambda,w)$. We find that it depends
on the value of the exponent $\mu$ as follows:

\noindent {\bf Case I: $\mu>2$}. Here one finds to leading order
in $(\lambda+w)$ that
\begin{equation}
Q(\lambda,w) = -\overline{ \left( \frac{1}{W} \right) }
(\lambda+w)+ \overline{ \left( \frac{1}{W^2} \right) }
\frac{(\lambda+w)^2}{2} \;. \label{eq:Qmu2}
\end{equation}

\noindent {\bf Case II: $1<\mu<2$}. Here $\overline{ W^{-2}}$
diverges and one has
\begin{equation}
Q(\lambda,w) = -\overline{ \left( \frac{1}{W} \right) }
(\lambda+w)+B (\lambda+w)^\mu \;,
 \label{eq:Qmu1}
\end{equation}
where $B=\overline{A} \pi / \mu \sin(\pi \mu)$. Here,
$\overline{A}$ is the amplitude of the tail of the probability
distribution, $\Psi(W) \approx \overline{A} W^{\mu-1}$.

\noindent {\bf Case III: $\mu<1$}. Here both $\overline{ W^{-1}}$
and $\overline{ W^{-2}}$ diverge and one has
\begin{equation}
Q(\lambda,w) =  \; B (\lambda+w)^\mu \label{eq:Qmu1} \;.
\end{equation}

With these results at hand we now turn to analyze the infinite
processivity limit.

\subsection{Infinite Processivity limit}

In this case we set $w=0$ for all sites. To analyze the eigenvalue
spectrum we rewrite Eq. (\ref{eq:toybc}) in the form
\begin{equation}
R(\lambda)=2 \pi i n/N \;. \label{eq:toybc2}
\end{equation}
Consider the solution for the different cases

\noindent {\bf Case I: $\mu>2$}. Here one has
\begin{equation}
2 \pi i n/N = -\overline{ \left( \frac{1}{W} \right) } \lambda+
\overline{ \left( \frac{1}{W^2} \right) } \frac{\lambda^2}{2} \;.
\end{equation}
It is straight forward to solve the quadratic equation for $\lambda$ and
realize that ${\rm Im}(\lambda) \propto n$ while ${\rm
Re}(\lambda) \propto n^2$ in agreement with the numerics of section~\ref{sect:inf-proc-spectr}.

\noindent {\bf Case II: $1<\mu<2$}. In this regime one has to
solve
\begin{equation}
2 \pi i n/N = -\overline{ \left( \frac{1}{W} \right) } (\lambda)+B
(\lambda)^\mu \;,
\end{equation}
where $B=\overline{A} \pi / \mu \sin(\pi \mu)$. Again, a
straightforward analysis shows that ${\rm Im}(\lambda) \propto n$
while ${\rm Re}(\lambda) \propto n^\mu$ in agreement with the
numerics.

\noindent {\bf Case III: $\mu<1$}. Finally, in this case one has
\begin{equation}
2 \pi i n/N =  \; B (\lambda)^\mu  \;,
\end{equation}
which evidently gives ${\rm Im}(\lambda) \propto n^{1/\mu}$ and
${\rm Re}(\lambda) \propto n^{1/\mu}$.

\subsection{Finite Processivity}

Next, we turn to the finite processivity case. Here it will be
sufficient to consider only the leading order contribution to
$R(\lambda)$.  To simplify, we study the case where $w$ can take
only two values $w_1=w$ and $w_2=0$, which occur with probability
$p_1$ and $p_2$. The general case can be analyzed very similarly.
This assumption gives, after performing the average over $w$,
\begin{equation}
R(E)=p_1 Q(E,w)+p_2Q(E,0)\;.
\end{equation}
As in the previous subsection we analyze behavior of the solutions
of the equations for $\lambda$ and $w$ small. Here it will only be
necessary to consider the cases $\mu>1$ and $\mu<1$.

\noindent {\bf Case I}: $\mu>1$. Here we have
\begin{equation}
-\langle \frac{1}{W} \rangle (p_1 (\lambda+w) + p_2 \lambda)= 2\pi
n i /N
\end{equation}
with $n$ an integer. Clearly, to solve the equation $\lambda$ must
have an imaginary part. (Note that we are missing the dependence
of the real part of the solution on $n$ since we neglected higher
order terms in $\lambda$). An imaginary part of the eigenvalue
implies that delocalized states exist for small disorder in $w$
\cite{HatanoNelson97}.

\noindent {\bf Case II}: $\mu<1$. Here one has
\begin{equation}
B(p_1 (\lambda+w)^\mu + p_2  \lambda^\mu)= \frac{2\pi n i}{N}
\label{eq:toyloc}
\end{equation}
Consider first the case $n=0$. In this case the eigenvalue must be
real \cite{VanKampen} and since the decay of particles from the
system can not be faster than $e^{-wt}$ one must have $-w \leq
\lambda <0$. Clearly, such a solution can not exist. Thus, the
assumption leading to Eq. (\ref{eq:toyloc}), that none of the
$P_k(\lambda)$ is zero, fails.  As we argued early,
$P_k(\lambda)=0$ away from a particular site, can only occur when
$\lambda$ is real. It is possible to show that in this case the
density is peaked near the site where $P_k(\lambda)=0$ decaying
exponentially fast as $k$ increases. We conclude that for any
strength of disorder in the hopping off rates the eigenfunctions
become localized. The lack of solution for cases with $n>0$ can be
proved by expanding around the lowest $| \lambda|$ (real)
eigenvalue and discovering that a solution is impossible.

{\bf Acknowledgments:} DKL and DRN thank Thomas Franosch for
helpful discussions during the early stages of this investigation.
DKL also benefitted from discussion with Piet Brouwer and
Christopher Mudry. Work by YK and DRN was supported by the
National Science Foundation through Grant DMR-0231631 and the
Harvard Materials Research Laboratory via Grant DMR-0213805. YK
was also supported through Grant DMR-0229243.

\appendix
\section{Simulations}
\label{apx:simualtions}

In the following we describe briefly the procedure we used to
simulate the model Eq. (\ref{eq:rates}). To make the simulation
efficient we first normalize the rates so that the largest one is
equal one. Then, at each step we choose with equal probability
between moving the motor to the right or left on the lattice.
Following this choice a random number is drawn from a uniform
distribution. The motor is moved in the chosen direction provided
the random number is smaller than the corresponding rate. This
protocol ensures relaxation to equilibrium in the absence of
chemical or mechanics driving forces. Time in the simulations is
measured by the number of attempted moves.

\section{Evolution Operator}
\label{ex:evolution}

Here we describe the construction of the evolution operator
corresponding to model Eq. (\ref{eq:rates}). First consider the
case when the rates for leaving the track are set to zero. To do
this we write the Master equation describing the time evolution as
\begin{equation}
\partial_t | P(t) \rangle = {\cal M}| P(t) \rangle \;,
\end{equation}
where $| P(t) \rangle$ is a vector with components $p_n(t)$, the
probability of being at site $n$ at time $t$ \cite{VanKampen}.
${\cal M}$ is the evolution operator of the model with non-zero
components given by
\begin{eqnarray}
{\cal M}_{i,i+1} &=& w_b^\leftarrow \;\;\;\;\;\; {\cal M}_{i,i-1}=
w_b^\rightarrow \;\;\; {\rm for} \; i \; {\rm even}
\nonumber \\
{\cal M}_{i,i+1} &=& w_a^\leftarrow \;\;\;\;\;\; {\cal M}_{i,i-1}=
w_a^\rightarrow \;\;\; {\rm for} \; i \; {\rm odd} \nonumber \\
{\cal M}_{i,i} &=& - {\cal M}_{i-1,i} - {\cal M}_{i+1,i} \;.
\label{eq:mat}
\end{eqnarray}
The last relation ensures conservation of probability. For a
heterogeneous model the rates corresponding to a given type of
monomer ${\cal M}_{i,i+1}$, $ {\cal M}_{i+1,i}$, $ {\cal
M}_{i+2,i+1}$, $ {\cal M}_{i+1,i+2}$ with $i$ even are chosen at
random. The diagonal terms are then automatically given by Eq.
\ref{eq:mat}. To study the eigenvalue spectrum periodic boundary
conditions are imposed. Finally, the hopping off rate from site
$i$, $h_i$ (given by $w^a_{\rm off}$ or $w^b_{\rm off}$ depending
on wether $i$ is even or odd) is added through ${\cal M}_{i,i} \to
{\cal M}_{i,i} - h_i$. The spectra presented in this paper were
calculated using \verb"MatLab".

\section{Random Energy Model}
In this Appendix we discuss the spectrum of the random energy
version of the model, described by Eq. \ref{eq:rates}, which also
allows the energy at even sites to vary. The rates are slightly
modified so that they to satisfy local detailed balance with
respect to these energies. An example of the relevance of random
energy models to biological system can be found in \cite{Slutsky},
which studies the diffusion of regulatory proteins along DNA.
Another realization (although there is no analogy of ``falling
off'') is translocation of ssDNA through a pore with identical
environments on the trans and cis sides \cite{Kafri04}. As stated
in the text, the long time large wavelength properties of this
model are very similar to those of a model studied in the context
of vortex physics \cite{HatanoNelson97}. We use model
(\ref{eq:rates}) with $T=1$, $\Delta \mu=0$, $T=1$ and use the
parameters $\{p_1\}=\{\alpha_1,\alpha_1',\omega_1,\omega_1',
\varepsilon_1, \varepsilon_1' \}=\{6,1,1,6,0,1.5\}$ and
$\{p_2\}=\{\alpha_2,\alpha_2',\omega_2,\omega_2',  \varepsilon_2,
\varepsilon_2'\}=\{1.2,1,1,1.2,0.5,0.8 \}$. Here $\varepsilon_i$,
$\varepsilon_i'$ denotes that energy at even and odd sites
respectively.

\begin{figure}
\includegraphics[width=8cm]{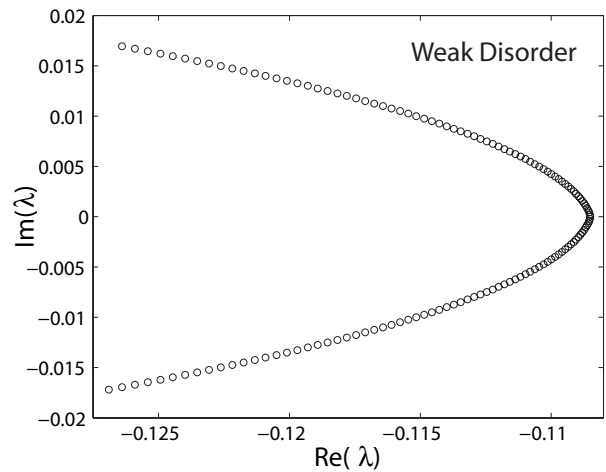} \caption{The eigenvalue spectrum for the random energy model with rates for hopping off the
track as specified in the text. Here an example of the weak
disorder regime spectrum is shown. Note that the right most part
of the ${\rm Re}(\lambda)$ axis is displaced slightly below zero.
Shown are the $140$ eigenvalues with the lowest value of
$|\lambda|$ and the system size is
$N=4500$.\label{fig:spectrumREL}}
\end{figure}

\begin{figure}
\includegraphics[width=8cm]{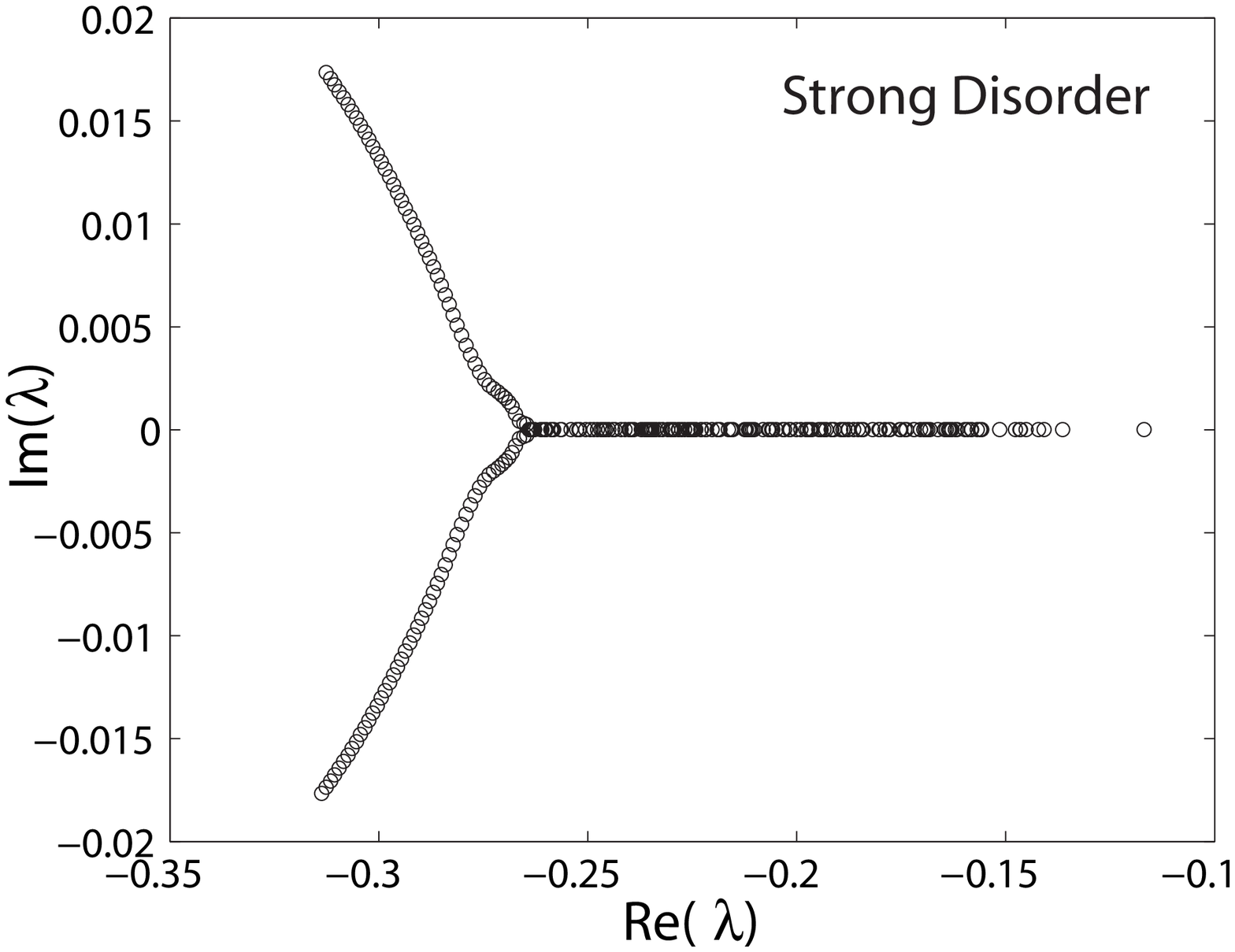} \caption{The eigenvalue spectrum for the random energy model with rates for hopping off the
track as specified in the text. Here an example of the strong
disorder regime spectrum is shown. Note that the right most part
of the ${\rm Re}(\lambda)$ axis is displaced slightly below zero.
Shown are the $300$ eigenvalues with the lowest value of
$|\lambda|$ and the system size is
$N=4500$.\label{fig:spectrumREH}}
\end{figure}

In Fig. \ref{fig:spectrumREL} and \ref{fig:spectrumREH} the
spectra of the evolution operator of the model are shown for weak
($w_{\rm off}^a=0.1$ and $w_{\rm off}^b=0.1$ ($w_{\rm off}^a=0.12$
and $w_{\rm off}^b=0.12$) for the parameter set $\lbrace p_1
\rbrace$ ($\lbrace p_2 \rbrace$)) and strong disorder ($w_{\rm
off}^a=0.1$ and $w_{\rm off}^b=0.1$ ($w_{\rm off}^a=0.4$ and
$w_{\rm off}^b=0.4$) for the set $\lbrace p_1 \rbrace$ ($\lbrace
p_2 \rbrace$) in the detachment rates. As can be seen from the
figures strong enough disorder in the hopping off rates, as
before, causes eigenvalue with small $| \lambda |$ to be real and
therefore associated with localized eigenfunctions
\cite{HatanoNelson97}.

\begin{thebibliography}{99}
\bibitem{com2} Note that the same energy landscape could be
obtained by decimating over even sites and allowing $\Delta
\varepsilon$ to vary.
\end{thebibliography}
\end{document}